# Fabrication of Bragg gratings in sub-wavelength diameter As$_2$Se$_3$ chalcogenide wires

Raja Ahmad[*], Martin Rochette and Chams Baker

*Department of Electrical and Computer Engineering, McGill University, Montreal (QC), Canada, H3A 2A7*
*Corresponding author: raja.ahmad@mail.mcgill.ca

The inscription of Bragg gratings in chalcogenide (As$_2$Se$_3$) wires with sub-wavelength diameter is proposed and demonstrated. A modified transverse holographic method employing He-Ne laser light at a wavelength of $\lambda_w$ = 633 nm allows the writing of Bragg grating reflectors in the 1550 nm band. The gratings reach an extinction ratio of 40 dB in transmission and a negative photo-induced index change of $\delta n \sim 10^{-2}$. The inscription of Bragg gratings in chalcogenide microwires will enable the fabrication of new devices with applications in nonlinear optics, and sensing in the near-to-mid-infrared region of wavelengths. © 2011 Optical Society of America

Over the last two decades, fiber Bragg gratings (FBGs) have emerged as one of the most widely used fiber optic devices. As linear devices, FBGs are used for sensing in various application fields ranging from bio/chemical systems to mechanical ones [1, 2]. FBGs have also received considerable scientific interest in nonlinear applications [3, 4] and have been utilized for all-optical switching [5, 6], pulse-shaping [7], enhancement of super-continuum generation [8] and for slowing down the speed of light [9]. However, the minimum peak power required to observe nonlinear effects in silica FBGs is in the order of 1 kW. The use of sub-wavelength diameter fibers or microwires, where the mode intensity is greatly increased and a considerable fraction of the mode power is present in the evanescent field outside the microwire, not only reduces the optical power to observe nonlinear effects in FBGs, but also enhances the sensitivity to the outside. The photo-inscription of Bragg gratings in silica microwires is however hard to achieve because the photosensitive core of the silica fiber vanishes upon tapering to such small diameters, which renders the wire insensitive to photo-inscription. Alternate techniques, such as by using femtosecond laser radiation [10], focused ion beam milling [11], metal deposition [12] and plasma etch postprocessing [13], have been utilized in past to fabricate Bragg gratings in microwires with diameters ranging from a few microns to tens of microns. However, these techniques are technologically challenging and often lead to surface damages. As of today, the photo-inscription of Bragg gratings in microwires is desirable, but has not been achieved so far.

Arsenic triselenide (As$_2$Se$_3$) is one of the chalcogenide glasses emerging as promising candidates for photonics and sensing applications [14]. As$_2$Se$_3$ is known to be highly photosensitive, it has a nonlinear coefficient that is ~ 930 times larger than that of silica glass and it is transparent in the 1-15 μm wavelength window [15]. As a result, the combination of microwire fabrication, the use of As$_2$Se$_3$ glass and the photo-inscription of FBGs is a promising approach for linear and nonlinear applications in the near-to-mid infrared region of wavelengths.

Recently, we have reported the fabrication of hybrid microwires made from an As$_2$Se$_3$ fiber surrounded by a protective PMMA coating [17] and observed a high waveguide nonlinear coefficient of $\gamma$ = 133 W$^{-1}$-m$^{-1}$. The PMMA coating provides mechanical strength for normal handling of the microwire and allows controlling the level of evanescent interaction with the surrounding environment. In this letter, we report the first inscription of Bragg gratings in chalcogenide microwires. This is also the first time that Bragg gratings are being reported in any optical fiber waveguide with sub-wavelength diameter. The transmission spectrum evolution as a function of time, both during and after the holographic exposure, is recorded and analyzed. A theoretical fit with the coupled mode theory is provided to reveal the grating parameters. Also, we observe that the refractive index of As$_2$Se$_3$ microwires decreases upon exposure to 633 nm light.

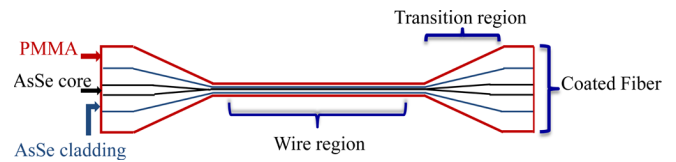

Fig. 1. Schematic depicting the various parts of a hybrid taper, including the (micro-) wire central section.

The chalcogenide fiber used for the experiment is provided by CorActive High-Tech inc. The fiber has a core/cladding diameter of 7/170 μm and a numerical aperture of 0.2. The fiber is coated with a protective layer of PMMA which is mostly transparent to the photo-inscription wavelength of $\lambda_w$ = 633 nm (absorption coefficient = 5.7 ×10$^{-4}$ cm$^{-1}$ [18]). The fiber is butt-coupled to a standard single-mode fiber made of silica and the two fibers are bonded permanently with UV epoxy. A hybrid As$_2$Se$_3$/PMMA microwire is then fabricated from an adiabatic tapering process as described in [17]. Fig. 1 depicts the various parts of a typical microwire. After tapering, the diameter and length of the As$_2$Se$_3$ wire region is 1 μm and 3 cm, respectively.

The grating fabrication setup consists of a modified Mach-Zehnder type interferometer, as shown in Fig. 2(a). A He-Ne laser source (Spectra Physics, Model 106-1) at a wavelength of $\lambda_w = 633$ nm provides the holographic photo-inscription pattern. The absorption coefficient of $As_2Se_3$ at this wavelength is $1.5 \times 10^4$ cm$^{-1}$ [19, 20]. This high absorption coefficient at the writing wavelength allows for a quick grating fabrication process. The output of the He-Ne laser source is split into two coherent beams that interfere at the inner surface of a glass prism and their angle $\theta$, as shown in Fig. 2(b), with respect to the prism surface is adjusted to achieve Bragg gratings with a first order resonance wavelength in the telecommunications C/L-band. The Bragg wavelength $\lambda_{Bragg}$ is controlled by the period $\Lambda$ of the holographic pattern, which depends on the angle $2\varphi$ between the two interfering beams inside the prism, as given below.

$$\lambda_{Bragg} = 2n_{eff}\Lambda \quad (1)$$

$$\Lambda = \frac{\lambda_w}{2\sin\varphi} \quad (2)$$

where $n_{eff}$ is the effective index of the propagating mode. The internal angle $\varphi$ is in turn defined by the external angle $\theta$, and the two angles in the case of a right-angled prism are related by the Eq. 3, as follows.

$$n_p \sin(\varphi - 45°) = \sin(\theta) \quad (3)$$

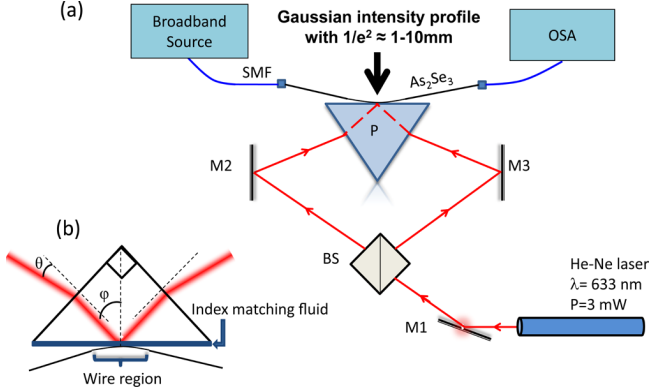

Fig. 3. Experimental setup for the Bragg grating photo-inscription and *in situ* monitoring of the process. SMF: single-mode fiber, P: prism, OSA: optical spectrum analyzer, M1, M2, and M3: reflecting mirrors, BS: beam-splitter. (b) detailed schematic of the prism where the microwire is placed during the grating growth, with the various angles defined in text, being labeled here.

The microwire is placed over the external surface of the prism with an index matching fluid, as depicted in Fig. 2(b), filling the gap between the prism and the microwire in order to maximize the transmission at the prism interface. The interfering beams are expanded using a focusing lens in each arm of the interferometer. The interference pattern has a Gaussian intensity profile with an adjustable $1/e^2$ full-width of 1-10 mm and a total writing power of 3 mW. The polarization of the two beams is identical and perpendicular with respect to the microwire axis in order to maximize the interference pattern contrast. The setup allows an *in-situ* monitoring of the grating growth process, with a broadband signal sent through the grating and observed on an optical spectrum analyzer.

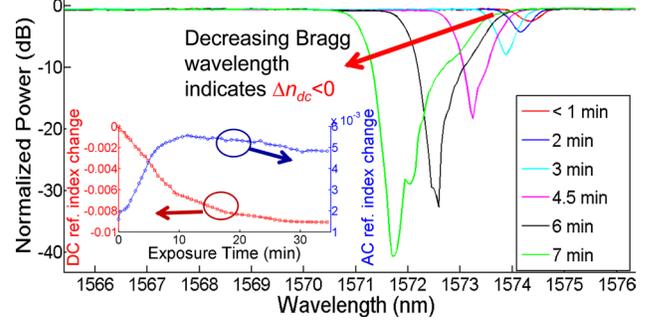

Fig. 2. Transmissivity of the Bragg grating as a function of time during photo-exposure, illustrating the grating growth dynamics. (Inset) Evolution of AC and DC refractive index change during the photo-exposure is also shown.

Fig. 3 shows the growth dynamics of the Bragg grating during the photo-exposure. The grating length in this case is 8 mm and the interference pattern is apodized. A reversible and wavelength independent transmission loss of 0.5 dB occurs during the process of photo-exposition, which can be observed from the spectra in Fig. 3. A dip in the transmission spectrum appears at $\lambda = 1574.4$ nm, and reaches -8 dB in less than 3 minutes of exposure and eventually, -40 dB after 7 minutes of exposure. Two observations can be made during this process: (1) the Bragg wavelength shifts towards shorter wavelengths, and (2) the width of the Bragg resonance increases. The first observation reveals that the refractive index of $As_2Se_3$ glass decreases upon the photo-exposure. This is supported by the appearance of grating apodization representative spectral features next to the longer wavelength edge of Bragg resonance (which is clearer in Fig. 4). Note that the previous studies on $As_2Se_3$ thin films have in contrast, reported an increase in refractive index upon exposure to 633 nm light [21, 22], which suggests that the refractive index change depends on the waveguide structure or the composition of the chalcogenide glass. A similar photoinduced index decrease in $As_2Se_3$ fiber was observed in an earlier report [23], but without quantification. The second observation during the photo-exposure reveals an increase in AC refractive index of the grating. These two observations lead us to quanitfy the changes in DC and AC refractive indices of the grating i.e., $\Delta n_{DC}(t)$ and $\Delta n_{AC}(t)$ respectively, by using the following relations.

$$\Delta n_{DC}(t) = n_0 \frac{\lambda_{B,current}(t) - \lambda_{B,initial}}{\lambda_{B,initial}} \quad (4)$$

$$\Delta n_{AC}(t) = n_0 \frac{\Delta\lambda_{B,current}(t)}{\lambda_{B,current}(t)} \quad (5)$$

where, $n_0$ (= 2.71) is the effective index of the microwire, calculated using the beam propagation method; $\lambda_{B,current}(t)$ and $\lambda_{B,initial}$, in eq. (4) are the current and initial Bragg wavelengths respectively; and the variable $\Delta\lambda_{B,current}(t)$, in eq. (5) is the width of the current Bragg resonance. The temporal evolution of $\Delta n_{DC}(t)$ and $\Delta n_{AC}(t)$ is shown as inset in Fig. 3. An AC refractive index change as high as $6.0 \times 10^{-3}$ is observed and a DC refractive index change of

$10^{-2}$ is observed. The AC refractive index increases to a maximum value after ~10 minutes of photo-exposure, and then decreases. This follows from a decrease in modulation depth of the holographic pattern. In fact, during the photo-exposure of the microwire, when the refractive index at the maxima of interference pattern decreases and eventually begins to saturate, the refractive index at the minima keeps decreasing at the same rate, thereby decreasing the modulation depth or the index contrast. This can be verified from Fig. 3 (inset), where it is shown that the $\Delta n_{AC}(t)$ starts dropping at the same time when the slope of $\Delta n_{DC}(t)$ decreases, which indicates the onset of saturation. We find that the DC refractive index continues to decrease even up to 4 hours of photo-exposure, although, the rate of change of $\Delta n_{DC}(t)$ becomes very small after the first 30 minutes of exposure.

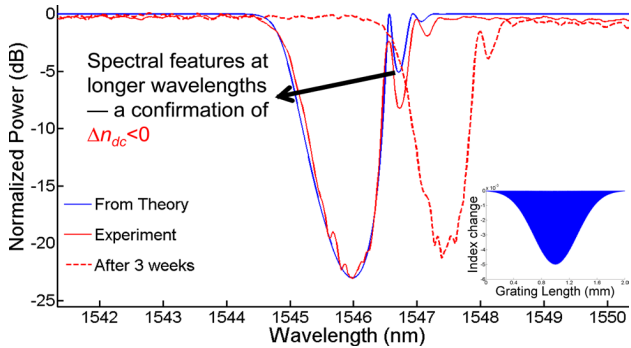

Fig. 4. Transmission spectrum of a 1mm long Bragg grating (red curve), and a theoretical fit of transmission spectrum with coupled mode theory (blue curve). Also, shown is the transmission spectrum of the grating after 3 weeks of exposure to ambient light. (Inset) Grating index profile used for the simulation, assuming the grating to be apodized following a gaussian profile.

Fig. 4 shows the spectrum of another grating, ~1 mm in length. A fit of the measured spectrum with coupled mode theory reveals a photoinduced refractive index change of $\Delta n_{AC} = 2.5 \times 10^{-3}$, which corresponds to a grating strength $\kappa L$ of 5.1 – $\kappa$ being the coupling coefficient given by $\pi \Delta n_{AC}/\lambda_{B,current}(t)$, and $L$ being the grating length. The presence of spectral features only on the longer wavelength side of the Bragg resonance shows that the grating is well-apodized. We also studied the aging of the grating at room temperature and observed a shift of Bragg wavelength by 1.5 nm towards the longer wavelengths, after 3 weeks of aging. The spectrum of the grating, as shown in Fig. 4, experiences no drastic degradation, which shows that the grating is quite stable in lab environment.

In conclusion, we have fabricated the first Bragg gratings in chalcogenide microwires with sub-wavelength diameters. The transmission spectrum shows a -8 dB dip at λ = 1574 nm within 3 minutes of exposure with a 3 mW laser interference pattern at a wavelength of 633 nm. The Bragg grating dip shifts to 1571.5 nm after 30 minutes, while growing to -40 dB. The observation of the transmission spectrum profile during exposure and subsequent 3 weeks of annealing reveal that the refractive index of $As_2Se_3$ decreases under exposition to 633 nm light and the grating strength remains stable. This device will find applications in sensing and nonlinear devices and for mid-infrared light processing.

R. A. gratefully acknowledges fruitful discussions with Prof. Suzanne Lacroix. This work was supported by the Fonds Québecois pour la Recherche sur la Nature et les Technologies (FQRNT).

### References

1. W. Liang, Y. Huang, Y. Xu, R.K. Lee, A. Yariv, Appl. Phys. Lett. **86**, 151122 (1-3) (2005).
2. A.D. Kersey, M.A. Davis, H.J. Patrick, M. LeBlanc, K.P. Koo, C.G. Askins, M.A. Putnam, E.J. Friebele, J. Lightwave Technol. **15**, 1442–1463 (1997).
3. B.J. Eggleton, R.E. Slusher, C.M. de Sterke, P.A. Krug, J.E. Sipe, Phys. Rev. Lett. **76**, 1627-1630 (1996).
4. H.G. Winful and V. Perlin, Phys. Rev. Lett. **84**, 3586-3589 (2000).
5. S. Larochelle, Y. Hibino, V. Mizrahi, G.I. Stegeman, Electron. Lett. **26**, 1459-1460 (1990).
6. I. V. Kabakova, D. Grobnic, S. Mihailov, E. C. Mägi, C. M. de Sterke, B.J. Eggleton, Opt. Express **19**, 5868-5873 (2011).
7. N. G. R. Broderick, D. Taverner, D. J. Richardson, M. Ibsen, R. I. Laming, Phys. Rev. Lett. **79**, 4566-4569 (1997).
8. P. S. Westbrook, J. W. Nicholson, K. S. Feder, Y. Li, T. Brown, Appl. Phys. Lett. **85**, 4600-4602 (2004).
9. J.T. Mok, C.M. de Sterke, I.C. M. Littler, B.J. Eggleton, Nat. Phys. **2**, 775-780 (2006).
10. X. Fang, C. R. Liao, D. N. Wang, Opt. Lett. **35**, 1007-1009 (2010).
11. V. Hodzic, J. Orloff, C.C. Davis, J. Lightwave Technol. **22**, 1610–1614, (2004).
12. W. Ding, S. R. Andrews, T. A. Birks, S. A. Maier, Opt. Lett. 31, 2556-2558 (2006).
13. W. Ding, S. R. Andrews, S. A. Maier, Opt. Lett. **32**, 2499-2501 (2007).
14. B. J. Eggleton, B. Luther-Davies, and K. Richardson, Nat. Photonics **5**(3), 141–148 (2011).
15. R.E. Slusher, G. Lenz, J. Hodelin, J. Sanghera, L. B. Shaw, I.D. Aggarwal, J. Opt. Soc. Am. B **21**, 1146-1155 (2004).
16. I. D. Aggarwal and J. S. Sanghera, J. Optoelectron. and Adv. Mat. **4**, 665 – 678 (2002).
17. C. Baker and M. Rochette, Opt. Express **18**, 12391-12398 (2010)
18. T. Kaino, K. Jinguji, S. Nara, Appl. Phys. Lett. **42**, 567 (1983).
19. J. P. De Neufville, S. C. Moss, S. R. Ovshinsky, J. Non-Cryst. Solids, **13**, 191-223 (1974).
20. J.B.R-Malo, E. Marquez, C. Corrales, P. Villares, R. Jimenez-Garay, Mat. Sci. Eng. B-Solid. **25**, 53-59 (1994).
21. A. van Popta, R. DeCorby, C. Haugen, T. Robinson, J. McMullin, D. Tonchev, S. Kasap, Opt. Express **10**, 639-644 (2002).
22. T.G. Robinson, R.G. DeCorby, J.N. McMullin, C.J. Haugen, S.O. Kasap, D. Tonchev, Opt. Lett. **28**, 459-461 (2003).
23. G.A. Brawley, V.G. Ta'eed, J.A. Bolger, J.S. Sanghera, I. Aggarwal, B.J. Eggleton, Electron. Lett. **44**, 846-847 (2008).

## References

1. W. Liang, Y. Huang, Y. Xu, R.K. Lee, and A. Yariv, "Highly sensitive fiber Bragg grating refractive index sensors," Appl. Phys. Lett. **86**, 151122 (1-3) (2005).
2. A.D. Kersey, M.A. Davis, H.J. Patrick, M. LeBlanc, K.P. Koo, C.G. Askins, M.A. Putnam, E.J. Friebele, "Fiber grating sensors," J. Lightwave Technol. **15**, 1442–1463 (1997).
3. B.J. Eggleton, R.E. Slusher, C.M. de Sterke, P.A. Krug, and J.E. Sipe, "Bragg grating solitons," Phys. Rev. Lett. **76**, 1627-1630 (1996).
4. H.G. Winful and V. Perlin, "Raman Gap Solitons," Phys. Rev. Lett. **84**, 3586-3589 (2000).
5. S. Larochelle, Y. Hibino, V. Mizrahi, and G.I. Stegeman, "All-optical switching of grating transmission using cross-phase modulation in optical fibers," Electron. Lett. **26**, 1459-1460 (1990).
6. Irina V. Kabakova, Dan Grobnic, Stephen Mihailov, Eric C. Mägi, C. Martijn de Sterke, and Benjamin J. Eggleton, "Bragg grating-based optical switching in a bismuth-oxide fiber with strong $\chi^{(3)}$-nonlinearity," Opt. Express **19**, 5868-5873 (2011).
7. N. G. R. Broderick, D. Taverner, D. J. Richardson, M. Ibsen, and R. I. Laming, "Optical Pulse Compression in Fiber Bragg Gratings," Phys. Rev. Lett. **79**, 4566-4569 (1997).
8. P. S. Westbrook, J. W. Nicholson, K. S. Feder, Y. Li, and T. Brown, "Supercontinuum generation in a fiber grating," Appl. Phys. Lett. **85**, 4600-4602 (2004).
9. J.T. Mok, C.M. de Sterke, I.C. M. Littler and B.J. Eggleton, "Dispersionless slow light using gap solitons," Nat. Phys. **2**, 775-780 (2006).
10. X. Fang, C. R. Liao, and D. N. Wang, "Femtosecond laser fabricated fiber Bragg grating in microfiber for refractive index sensing," Opt. Lett. **35**, 1007-1009 (2010).
11. V. Hodzic, J. Orloff, and C.C. Davis, "Periodic structures on biconically tapered optical fibers using ion beam milling and boron implantation," J. Lightwave Technol. **22**, 1610–1614, (2004).
12. W. Ding, S. R. Andrews, T. A. Birks, and S. A. Maier, "Modal coupling in fiber tapers decorated with metallic surface gratings," Opt. Lett. 31, 2556-2558 (2006).
13. W. Ding, S. R. Andrews, and S. A. Maier, "Surface corrugation Bragg gratings on optical fiber tapers created via plasma etch postprocessing," Opt. Lett. **32**, 2499-2501 (2007).
14. B. J. Eggleton, B. Luther-Davies, and K. Richardson, "Chalcogenide Photonics," Nat. Photonics **5**(3), 141–148 (2011).
15. Richart E. Slusher, Gadi Lenz, Juan Hodelin, Jasbinder Sanghera, L. Brandon Shaw, and Ishwar D. Aggarwal, "Large Raman gain and nonlinear phase shifts in high-purity $As_2Se_3$ chalcogenide fibers," J. Opt. Soc. Am. B **21**, 1146-1155 (2004).
16. I. D. Aggarwal and J. S. Sanghera, "Development and applications of chalcogenide glass optical fibers at NRL", Journal of Optoelectronics and Advanced Materials, **4** (3), September 2002, p. 665 – 678.
17. Chams Baker and Martin Rochette, "Highly nonlinear hybrid AsSe-PMMA microtapers," Opt. Express **18**, 12391-12398 (2010)
18. T. Kaino, K. Jinguji, and S. Nara, "Low loss poly (methylmethacrylate-d8) core optical fibers," Appl. Phys. Lett. **42**, 567 (1983).
19. J. P. De Neufville, S. C. Moss, and S. R. Ovshinsky, "Photostructural transformations in amorphous $As_2Se_3$ and $As_2S_3$ films," J. Non-Cryst. Solids, **13**, 191-223 (1974).
20. J.B. Ramirez-Malo, E. Marquez, C. Corrales, P. Villares, and R. Jimenez-Garay, "Optical characterization of $As_2S_3$ and $As_2Se_3$ semiconducting glass films of non-uniform thickness from transmission measurements," Mat. Sci. Eng. B-Solid. **25**, 53-59 (1994).
21. A. van Popta, R. DeCorby, C. Haugen, T. Robinson, J. McMullin, D. Tonchev, and S. Kasap, "Photoinduced refractive index change in As2Se3 by 633nm illumination," Opt. Express **10**, 639-644 (2002)
22. Travis G. Robinson, Ray G. DeCorby, James N. McMullin, Chris J. Haugen, Safa O. Kasap, and Dancho Tonchev, "Strong Bragg gratings photoinduced by 633-nm illumination in evaporated $As_2Se_3$ thin films," Opt. Lett. **28**, 459-461 (2003).
23. G.A. Brawley, V.G. Ta'eed, J.A. Bolger, J.S. Sanghera, I. Aggarwal, B.J. Eggleton, "Strong photoinduced Bragg gratings in arsenic selenide optical fibre using transverse holographic method," Electron. Lett. **44**, 846-847 (2008).